\documentclass[letter]{aa}

\usepackage{natbib}
\usepackage{graphicx}
\usepackage{txfonts}
\usepackage{color}
\usepackage{multirow}

\newcommand{\hso}{\emph{Herschel} Space Observatory}
\newcommand{\herschel}{\emph{Herschel}}

\newcommand{\hcl}{HC$\ell$}
\newcommand{\cl}{C$\ell$}
\newcommand{\hbrp}{HBr$^{+}$}

\begin{document}

\title{Interstellar bromine abundance is consistent with cometary ices from \emph{Rosetta}}

\author{
        N.F.W. Ligterink\inst{1,2}
        \and
        M. Kama\inst{3,1}
}

   \institute{
             Leiden Observatory, Leiden University, PO Box 9513, 2300 RA Leiden, The Netherlands\\     
             \email{ligterink@strw.leidenuniv.nl}\\     
             \and
             Raymond and Beverly Sackler Laboratory for Astrophysics, Leiden Observatory, Leiden University, PO Box 9513, 2300 RA Leiden, The Netherlands\\ \and
             Institute of Astronomy, University of Cambridge, Madingley Road, Cambridge, CB3 0HA, United Kingdom\\                     
             \email{mkama@ast.cam.ac.uk}\\ 
             }

\date{}

\abstract
{Cometary ices are formed during star and planet formation, and their molecular and elemental makeup can be related to the early solar system via the study of inter- and protostellar material.}
{We set out to place the first observational constraints on the interstellar gas-phase abundance of bromine (Br). We further aim to compare the protostellar Br abundance with that measured by \emph{Rosetta} in the ices of comet 67P/Churyumov-Gerasimenko.}
{Archival \herschel\ data of Orion~KL, Sgr~B2(N), and NGC~6334I are examined for the presence of HBr and \hbrp\ emission or absorption lines. A chemical network for modelling HBr in protostellar molecular gas is compiled to aid in the interpretation.}
{HBr and \hbrp\ were not detected towards any of our targets. However, in the Orion~KL Hot Core, our upper limit on HBr/H$_{2}$O is a factor of ten below the ratio measured in comet 67P. This result is consistent with the chemical network prediction that HBr is not a dominant gas-phase Br carrier. Cometary HBr is likely predominantly formed in icy grain mantles which lock up nearly all elemental Br.}
{}

   \maketitle
   

\section{Introduction}

Observations of halogen-bearing species in molecular gas can probe the gas-to-ice depletion of volatile elements during star and planet formation \citep{Gerinetal2016}. Previous studies have characterized the abundance of fluorine (F) and chlorine (\cl) in protostellar gas.  We analyse archival data from the \hso\ to constrain the gas-phase abundance budget of bromine (Br) and thus expand the overall knowledge of interstellar halogen chemistry. With the recent detection of the organohalogen CH$_{3}$\cl\ \citep{Fayolle2017}, constraints on the abundances of Br could also give information on the presence of organobromine compounds in the interstellar medium.

The solar abundances of F and \cl\ are $(3.63\pm0.11)\times10^{-8}$ and $(3.16\pm0.95)\times10^{-7}$, respectively \citep{Asplundetal2009}. The gas-phase \hcl\ abundance in dense protostellar cores is [\hcl/H$_{2}$]${\sim}10^{-10}$, with \cl\ depleted by a factor $100$--$1000$ \citep{Blakeetal1985, Schilkeetal1995, Zmuidzinasetal1995, Salezetal1996, NeufeldGreen1994, Pengetal2010, Kamaetal2015}. Models indicate that the missing \cl\ is in \hcl\ ice \citep{Kamaetal2015}. A high \cl\ fraction in \hcl\ ice was confirmed \emph{in-situ} for comet 67P/Churyumov-Gerasimenko (hereafter 67P/C-G) with \emph{Rosetta}, which recently measured \hcl/$\rm H_{2}O\approx 1.2\cdot10^{-4}$ \citep{Dhoogheetal2017}, close to \emph{Herschel} upper limits at comets Hartley 2 and Garradd \citep{BockeleeMorvanetal2014}. 

In contrast to F and \cl, the solar and interstellar Br abundance is unknown, but in meteorites it is equivalent to Br/H$=(3.47\pm0.02)\times 10^{-10}$ \citep{Loddersetal2009}. The two stable isotopes of bromine are $^{79}$Br and $^{81}$Br, with a terrestrial abundance ratio of $^{79}$Br/$^{81}$Br${=}1.03$ \citep{Boehlkeetal2005}. For comet 67P/C-G, the \emph{Rosetta} spacecraft detected HBr and measured an elemental ratio of Br/O~$=(1-7)\times10^{-6}$ in the inner coma, consistent with nearly all bromine being locked in ice, analogously to chlorine.

Accounting for the range of variation seen in 67P/C-G and the uncertainties in terrestrial data, the cometary Br/\cl\ value of $\approx0.02$ \citep{Dhoogheetal2017} is consistent with the bulk Earth estimate of Br/\cl~$\approx0.04$ \citep{Allegreetal2001}.

If Br has a similar depletion level as \cl\ in protostellar gas, it may be detectable as HBr at a sensitivity of $\delta T\lesssim 0.01\times T$(\hcl), where $T$ is the intensity in kelvin. The lowest rotational transitions of HBr are at around $500$, $1000$, and $1500\,$GHz. These frequencies are not observable from the ground, but were covered by the \emph{Herschel}/HIFI spectrometer. We also consider the potentially abundant molecular ion HBr$^{+}$. During regular science observations, HIFI serendipitously covered transitions of HBr and HBr$^{+}$ towards the bright protostellar regions Orion~KL, Sagittarius~B2 North (hereafter Sgr~B2(N)), and NGC6334I. We use these data to constrain the gas-phase abundance of Br-carriers. 

In Section~\ref{sec:data}, we summarise the spectroscopy and the archival \herschel\ data, which are analysed and discussed in Section~\ref{sec:results}. Section~\ref{sec:comp} compares the interstellar observations with cometary detections in 67P/C-G. In Section~\ref{sec:chem}, we review the interstellar chemistry of Br. Our conclusions are presented in Section~\ref{sec:conclusions}.

\section{Data}\label{sec:data}

\subsection{Spectroscopy of HBr and \hbrp}\label{sec:spec}

Measurements on rotational lines of HBr were performed by \citet{VanDijkDymanus1969} for the hyperfine components of the first rotational transition and later extended by \citet{DiLonardoetal1991} up to $J_{\rm u}=9$. The first three rotational transitions of both H$^{79}$Br and H$^{81}$Br are found at frequencies just above $500$, $1000$, and $1500\,$GHz, respectively (Table \ref{tab:hbrspec} in the Appendix). The three lowest rotational transitions of \hbrp\ are found at $1188.2$, $1662.7,$ and $2136.8\,$GHz and also display hyperfine splitting \citep{SaykallyEvenson1979, Lubicetal1989}. However, insufficient spectroscopic data of \hbrp\ are available to determine column densities. The lowest HBr and \hbrp\ transition frequencies fall in spectral regions with heavy atmospheric absorption and are best observed from space.

\subsection{Archival $\emph{Herschel}$ observations and selected sources}

The \hso\ mission \citep{Pilbrattetal2010}, active from 2009 to 2013, was the most sensitive observatory to date in the terahertz frequency range. We investigate archival high spectral resolution and broad wavelength coverage data from its heterodyne instrument, HIFI \citep{deGraauwetal2010}.

The HEXOS guaranteed-time key program \citep[PI E.A. Bergin,][]{Berginetal2010} obtained full spectral scans of Orion~KL and Sgr~B2 \citep{Crockett2014b, Crockett2014a, Neill2014}, covering three rotational transitions of HBr in HIFI bands 1a, 4a and 6a and \hbrp\ transitions in band 5a and 6b. The CHESS key program \citep[PI C. Ceccarelli,][]{Ceccarelli2010} observed NGC6334I in the same HIFI bands \citep{Zernickeletal2012}. These three sources are bright and well-studied, and have yielded strong detections of the halogens HF and \hcl\, with the latter having integrated intensities of $\int T_{mb}dv$ = 701.9 K km s$^{-1}$ over two lines for Orion~KL and $\int T_{mb}dv$ = 40 K km s$^{-1}$ over three lines for NGC6334I. Line intensities one to two orders of magnitude lower than the \hcl\ peak brightness should be detectable, based on 3$\sigma$ noise levels of $\sim$0.36 K km s$^{-1}$ and $\sim$0.08 K km s$^{-1}$ for these sources, respectively. The observational details of these three sources are listed in Table~\ref{tab:3sig}.

\begin{table*}
      \caption[]{Source parameters in HIFI band 1a}
         \label{tab:3sig}
         \centering
         \begin{tabular}{l c c c c c c c}
            \hline
            \hline
            Source   & $\theta_{S}$ & RMS & $V_{\rm LSR}$ & $\Delta$V & 3$\sigma$ Flux & Continuum & Mean T$_{\rm ex}$  \\
                 & $^{\arcsec}$ & (mK) & ($\,$km$\,$s$^{-1}$) & ($\,$km$\,$s$^{-1}$) & (mK$\,$km$\,$s$^{-1}$) & K  & K\\
            \hline
            \noalign{\smallskip}
            Orion KL$^{a}$ Plat. & 30 & $23.0$ & 7 -- 11 & $\geq$20 & 364 & 1.6 & 94\\
            Orion KL HC & 10 & $23.0$ & 4 -- 6 & 7 & 216 & 1.6 & 155\\
            Orion KL CR & 10 & $23.0$ & 7 -- 9 & 6 & 200 & 1.6 & 138\\
            \noalign{\smallskip}
            Sgr B2(N)$^{b}$ HC & $\sim$5 & $29.3$ & 50 -- 100 & 9 & 311 & 2.5 & 144\\
            Sgr B2(N) env. & -- & $29.3$ & 50 -- 100 & 20 & 464 & 2.5 & 34\\
            \noalign{\smallskip}
            NGC6334I$^{c}$ HC & $\sim$5 & $7.8$ & -20 -- 7 & 4 & \phantom{-}55 & 1.0 & 90\\
            NGC6334I env. & 20* & $7.8$ & -20 -- 7 & 8 & \phantom{-}78 & 1.0 & 22\\
            \hline
         \end{tabular} 
          \flushleft
        \emph{Notes.} Plat. -- Plateau, HC -- Hot Core, CR -- Compact Ridge, env. -- envelope. $^{a}$\citet{Crockett2014b}; $^{b}$\citet{Neill2014}; $^{c}$\citet{Zernickel2012}; *based on the derived \hcl\ source size. Mean excitation temperatures are derived from detected species in publications $^{a,b,c}$.
\end{table*} 

\subsection{Analysis method}\label{method}

All sources are inspected for features corresponding to transitions of H$^{79/81}$Br and HBr$^{+}$ using the \texttt{Weeds} addition \citep{Maret2011} of the Continuum and Line Analysis Single-dish Software (CLASS\footnote{\texttt{http://www.iram.fr/IRAMFR/GILDAS}}). For line identification, we use the JPL\footnote{\texttt{http://spec.jpl.nasa.gov}} \citep{Pickett1998} and CDMS\footnote{\texttt{http://www.astro.uni-koeln.de/cdms}} \citep{Muller2001,Muller2005} spectroscopy databases. Source velocities matching previous detections of halogen-bearing molecules are considered most relevant, but we explore a large $V_{\rm LSR}$ range to check for emission or absorption components matching the hyperfine pattern. For emission features, the total column density $N_{\rm T}$ of a species can be calculated by assuming local thermodynamic equilibrium (LTE):
        \begin{eqnarray}
        \label{eq.col_em}
        \frac{3k_{\rm B}\int T_{\rm{MB}}dV}{8\pi^{3}\nu\mu^{2}S} = \frac{N_{\rm{up}}}{g_{\rm{up}}} = \frac{N_{\rm T}}{Q(T_{\rm{rot}})}e^{-E_{\rm{up}}/T_{\rm{rot}}},
        \end{eqnarray} 
   \\

where $\int$ $T_{\rm{MB}}$dV is the integrated main-beam intensity of a spectral line, $\nu$ the transition frequency, $\mu^{2}$ the dipole moment, $S$ the transition strength, $g_{\rm{up}}$ the upper state degeneracy, $Q(T_{\rm{rot}}$) the rotational partition function, $E_{\rm{up}}$ the upper state energy and $T_{\rm{rot}}$ the rotational temperature. Upper limits are given at 3$\sigma$ confidence and calculated by $\sigma = 1.1\sqrt{\delta \nu \Delta V}\cdot$RMS, where $\delta\nu$ is the velocity resolution, $\Delta V$ the line width (estimated based on other transitions in the spectrum) and RMS the root mean square noise in Kelvin. A factor of $1.1$ accounts for the flux calibration uncertainty of 10\% \citep{Roelfsemaetal2012}.

In the source sample, the hydrogen halides HF and \hcl\ are also found in absorption. We calculate the column density corresponding to absorption features from:
        \begin{eqnarray}
        \label{eq.col_ab}
        \tau = -ln\ \Bigg(\frac{T_{\rm MB}}{T_{\rm cont}}\Bigg)
        ,\end{eqnarray} 
   \\
   and
        \begin{eqnarray}
        \label{eq.col_ab}
        N_{\rm T} = \frac{8\pi^{3/2}\cdot\Delta V}{2\sqrt{ln2}\cdot\lambda^{3}} \frac{g_{\rm l}}{g_{\rm u}}\cdot\tau ,
        \end{eqnarray} 
   \\
where $\tau$ is the optical depth, $T_{\rm MB}$ the brightness temperature of the feature and $T_{\rm cont}$ the continuum level. $\lambda$ is the wavelength of the transitions and $g_{\rm l}$ and $g_{\rm u}$ are its lower and upper state degeneracies. For non-detections, a 3$\sigma$ upper limit column density is determined using $T_{\rm MB}$ = $T_{\rm cont}$-3$\cdot$RMS and assuming $\Delta V$ equals the average line width for other species in the source.

If the source does not fill the entire HIFI beam (at $500\,$GHz, $\theta_{\rm B}{=}44^{\arcsec}$), we correct the column densities for beam dilution by applying the factor $\eta_{\rm BF} = {\theta_{\rm S}^{2}}/(\theta_{\rm S}^{2}+\theta_{\rm B}^{2})$, where $\theta_{\rm S}$ is the source size and $\theta_{\rm B}$ the beam size. Source sizes are taken from literature, see Table \ref{tab:3sig}, and are used as the physical size of the emitting regions. Deviations from the actual emitting area of a species may occur and would result in different column densities. The \textit{source-}averaged column density is calculated from $N_{\rm S} = {N_{\rm T}}/{\eta_{\rm BF}}$. 

We determined the upper limit column densities from the HBr J = 1$_{x}$ $\rightarrow$ 0$_{2}$ transitions at $500\,$GHz in HIFI band 1a, because of the low noise levels in this frequency range. HBr is considered as the sum of its isotopologs, H$^{79}$Br and H$^{81}$Br. We assume that the cosmic and local isotope ratios are equal (in the solar system [$^{79}$Br]/[$^{81}$Br]${=}1.03$). Aside from the molecular mass, several spectroscopic parameters for transitions of both isotopes, such as $A_{\rm ij}$ and $E_{\rm up}$, are identical. The 1$_{3}$ $\rightarrow$ 0$_{2}$ line is the strongest hyperfine component and constrains the column density the most and is therefore used to give the most stringent upper limits.

\begin{figure}[h!]
  \centering
    \includegraphics[width=0.5\textwidth]{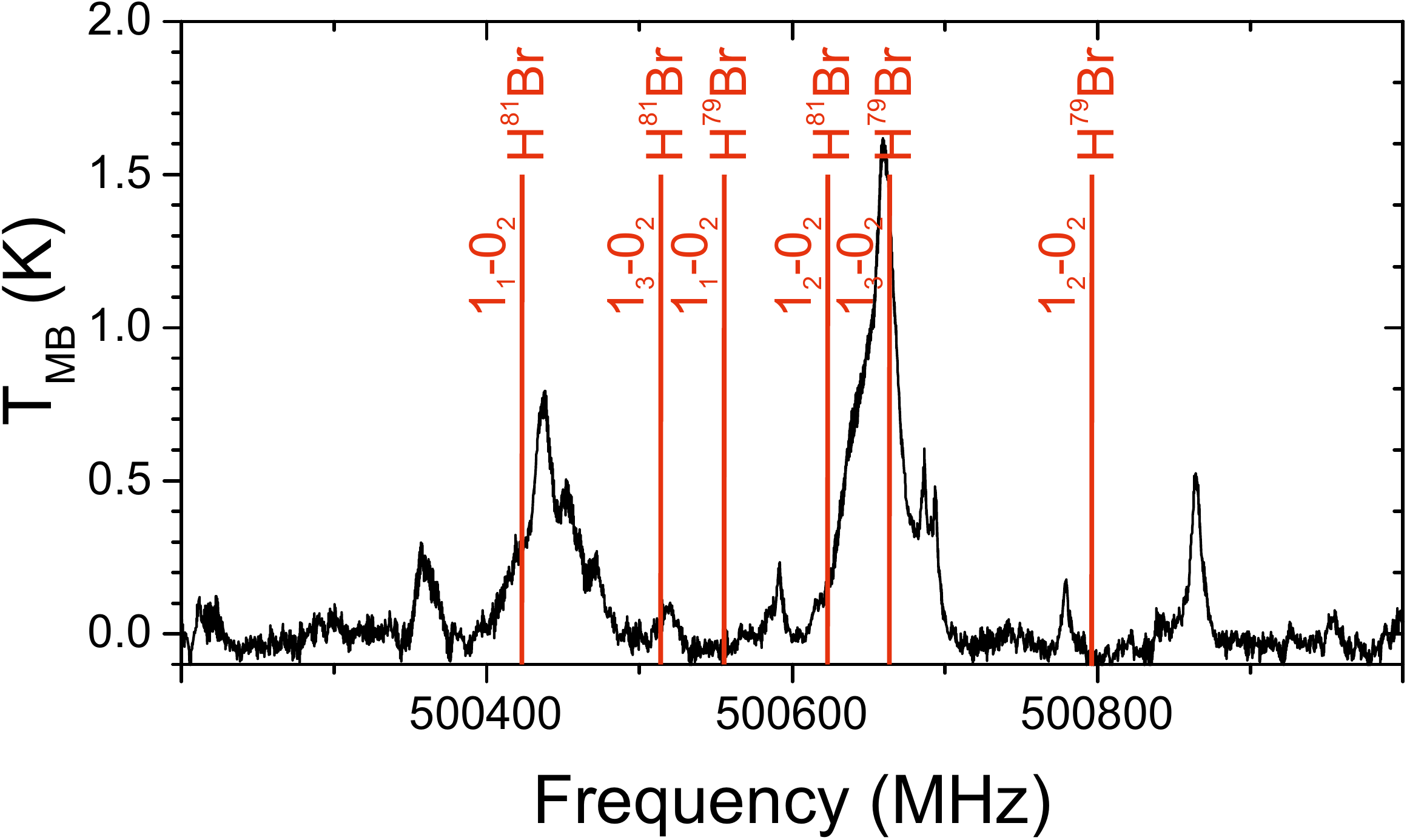}
    \caption{Positions of the J = 1$_{x}$ $\rightarrow$ 0$_{2}$ transitions of H$^{79}$Br and H$^{81}$Br at 500 GHz towards Orion KL for $V_{\rm LSR}$ = 9 $\,$km$\,$s$^{-1}$.}
    \label{fig:orion}
\end{figure}
%
%
\section{Search for HBr and \hbrp\ in the \herschel\ spectra}\label{sec:results}

Analysis of the HIFI spectra of Orion~KL, Sgr~B2(N), and NGC6334I yielded no detections of HBr or \hbrp\ features in emission or absorption. Figure~\ref{fig:orion} shows the positions of the HBr transitions at 500 GHz in the Orion KL spectrum at $V_{\rm LSR}$ = 9 $\,$km$\,$s$^{-1}$, corresponding to the average velocity of the Plateau components. The data were analysed over a large range of source velocities, mainly focussing on velocities of known components. For Sgr B2(N) and NGC6334I the same figures can be found in Appendix~\ref{ap.sources}. 

\section{Protostellar \emph{versus} cometary abundance}
\label{sec:comp}

The upper limit abundance ratios of HBr toward the protostellar sources can be compared with measurements taken by the \emph{Rosetta} mission of the coma gas of comet 67P/C-G \citep{Dhoogheetal2017}. We look at column density ratios of HBr (Table~\ref{tab:hbrcoldens}) with respect to those of H$_{2}$, H$_{2}$O, CH$_{3}$OH, HF, and \hcl\ (Table~\ref{tab:coldens}). The $N$(H$^{79+81}$Br)/$N$(X) column density ratios based either on emission or absorption upper limits for Orion~KL, Sgr~B2(N), and NGC6334I are listed in Table~\ref{tab:ratio}.

The upper limits in emission are based on an excitation temperature of 100~K, which is chosen to be within a factor of a few of all the detected molecules we compare with. For an assessment of the impact of $T_{\rm ex}$, Fig.~\ref{fig:rotTplot} shows the temperature dependence of the $3\,\sigma$ upper limit for the first three hyperfine transitions of H$^{79/81}$Br, including beam dilution correction for the Orion KL Hot Core. For Sgr~B2(N) and NGC6334I, a distinction is made between the hot core and envelope components. If a source contains multiple kinematic components of a species, we adopt the dominant one.

For 67P/C-G, \citet{Dhoogheetal2017} give the Br/O ratio and the CH$_{3}$OH abundance. The cometary halogens are equal to the halides (HX), but the O abundance is the sum of H$_{2}$O, CO, CO$_{2}$ and O$_{2}$. For the comet, we can therefore take Br$\equiv$HBr, and we further assume O$\approx$H$_{2}$O. A ratio of CH$_{3}$OH/H$_{2}$O = 3.1--5.5$\times$10$^{-3}$ has been measured by \citet{LeRoy2015}.

\begin{figure}[!ht]
\includegraphics[clip=,width=1.0\linewidth]{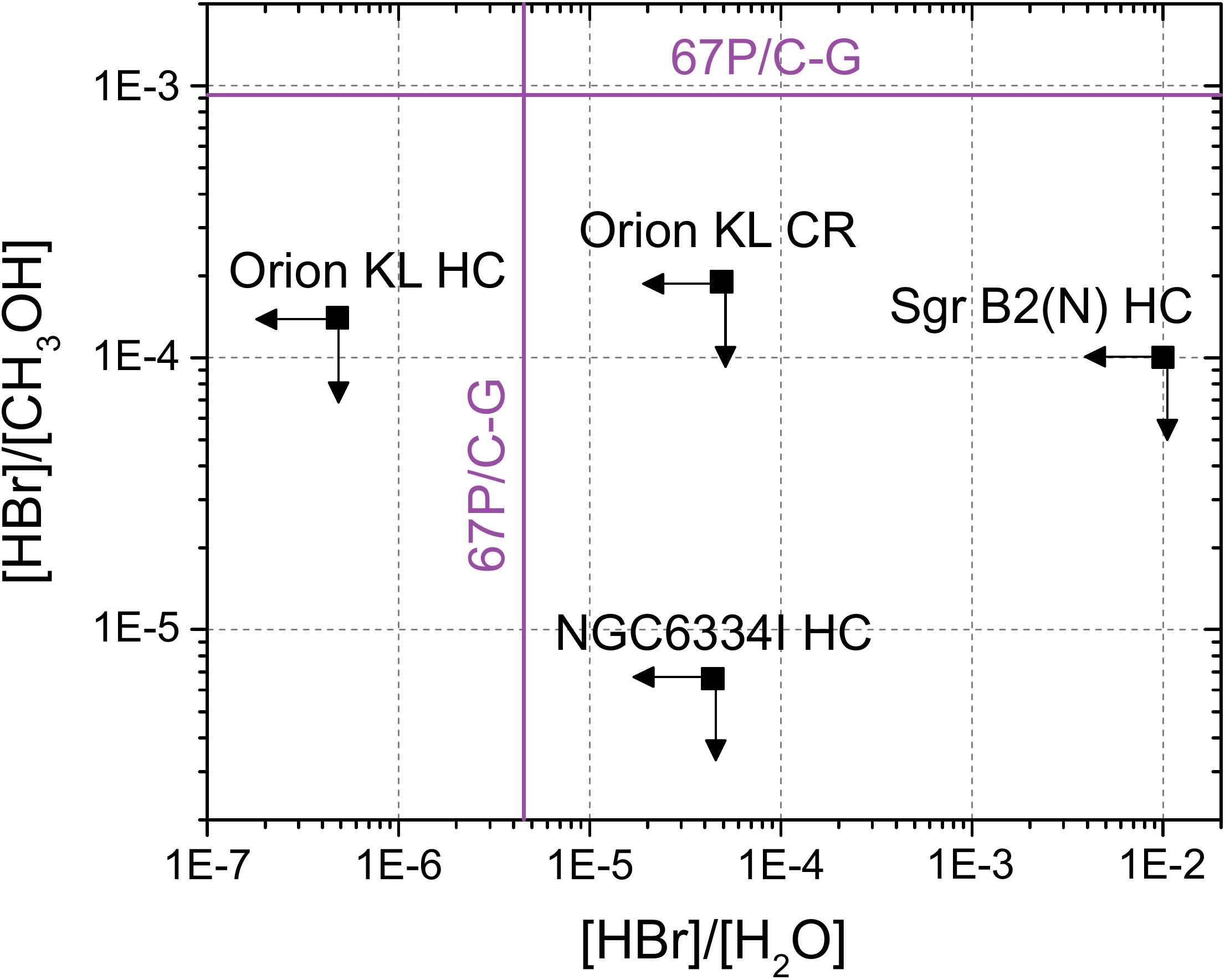}
\caption{The (H)Br/CH$_{3}$OH and (H)Br/(H$_{2}$)O ratios plotted for 67P/C-G \citep[purple lines,][]{Dhoogheetal2017} and the upper limits on these ratios for the protostar sample (this work).}
\label{fig:meoh_h2o_abundance}
\end{figure}

A comparison of HBr with H$_{2}$O and CH$_{3}$OH is shown in Fig.~\ref{fig:meoh_h2o_abundance}, and the full set of abundance ratios of HBr with other molecules is given in Table~\ref{tab:ratio}. The HBr/CH$_{3}$OH ratio in all our targets is constrained to be below that in comet 67P/C-G. This is not necessarily due to a particularly high methanol abundance in our targets, but rather could signify a low fraction of Br atoms locked up in gas-phase HBr molecules. The only source where we can constrain the HBr/H$_{2}$O ratio to be below that in 67P/C-G is the Orion~KL Hot Core. This may, again, be explained with a low fraction of Br atoms in gas-phase HBr. If all elemental bromine were in gaseous HBr, we would have expected to have made a detection in the Orion~KL Hot Core. A comparison with the cometary measurements suggests, then, that the HBr molecules are formed in icy grain mantles, rather than in the gas phase, or sublimate from the grain surface at a temperature higher than water.

The HBr/\hcl\ abundance ratio in the Orion~KL Plateau is constrained to be a factor $\gtrsim 4$ below that in 67P/C-G. The difficulties in forming a large abundance of HBr in the gas phase when \hcl\ is clearly present lead us to conclude that cometary HBr has an origin in grain surface chemistry in volatile-rich ice mantles. 

\begin{table*}
      \caption[]{Abundance ratios of H$^{79+81}$Br ($\equiv$B) upper limit column density with H$_{2}$, CH$_{3}$OH, H$_{2}$O, HF, and H$^{35+37}$\cl ($\equiv$\hcl) for Orion KL, Sgr B2(N), and NGC6334I, compared with abundance ratios derived for 67P/C-G.}
         \label{tab:ratio}
         \centering
         \begin{tabular}{l c c c c c}
            \hline
            \hline
            \noalign{\smallskip}
            Source & B/H$_{2}$ & B/H$_{2}$O & B/CH$_{3}$OH & B/HF & B/\hcl\ \\
            \hline
            \noalign{\smallskip}
            Orion KL Plat.$^{a}$ & $\leq$8.9$\times$10$^{-11}$ & $\leq$1.9$\times$10$^{-5}$ & -- & $\leq$1.2$\times$10$^{1}$* & $\leq$1.3$\times$10$^{-2}$ \\
            Orion KL HC$^{a}$  & $\leq$3.1$\times$10$^{-10}$ & $\leq$4.8$\times$10$^{-7}$ & $\leq$1.4$\times$10$^{-4}$ & -- & --  \\
            Orion KL CR$^{a}$ & $\leq$2.3$\times$10$^{-10}$ & $\leq$4.9$\times$10$^{-5}$ & $\leq$1.9$\times$10$^{-4}$ & -- & --  \\
            \noalign{\smallskip}
            Sgr B2(N) HC$^{b}$ & $\leq$6.5$\times$10$^{-11}$ & $\leq$1.0$\times$10$^{-2}$ & $\leq$1.0$\times$10$^{-4}$ & -- & -- \\
            Sgr B2(N) env.$^{b}$ & -- & -- & $\leq$1$\times$10$^{-3}$ & $\leq$1.1$\times$10$^{-1}$* & $\leq$6.9$\times$10$^{-2}$* \\
            \noalign{\smallskip}
            NGC6334I HC$^{c}$  & $\leq$8.4$\times$10$^{-11}$ & $\leq$4.4$\times$10$^{-5}$ & $\leq$6.6$\times$10$^{-6}$ & -- & $\leq$4.8$\times$10$^{-1}$ \\
            NGC6334I env.$^{c}$  & -- & -- & -- & $\leq$2.0$\times$10$^{-1}$* & $\leq$9.3$\times$10$^{-1}$*   \\
            \noalign{\smallskip}
            \hline
            \noalign{\smallskip}
            67P/C-G$^{a}$                       & -- & 4.5$^{+3.5}_{-3.5}\times$10$^{-6}$ $^{b}$ & 1.4$^{+1.2}_{-0.7}\times$10$^{-3}$ & 1.7$^{+5.1}_{-1.3}\times$10$^{-2}$ & 4.8$^{+13.8}_{-3.7}\times$10$^{-2}$ \\        
            \noalign{\smallskip}
            \hline
         \end{tabular} 
          
        \tablefoot{Plat. -- Plateau, HC -- Hot Core, CR -- Compact Ridge, env. -- envelope. $^{a}$\citet{Dhoogheetal2017}; $^{b}$Br/O elemental ratio, O has contributions of water, but also CO, CO$_{2}$ and O$_{2}$; *indicates values based on an upper limit in absorption, other values are based on the emission upper limits.}
\end{table*} 

\section{Interstellar chemistry of Br}\label{sec:chem}
The inter- and protostellar chemistry of bromine is poorly characterized, compared to that of fluorine and chlorine \citep[e.g.][]{Jura1974, Blakeetal1986, Schilkeetal1995, NeufeldWolfire2009}. In Table~\ref{tab:reactions}, we present a network compiled from published measurements and calculations, with missing data filled in with values from the \cl\ and F networks. Some reactions are not listed in this table, for these reactions we adopt the equivalent \cl\ reaction parameters of \citet[][their Table 1]{NeufeldWolfire2009}.

The neutral-neutral chemistry, reactions (1) to (3), is relatively well studied. The Br$+$H$_{2}$ reaction leading to HBr$+$H, with an $8812\,$K activation energy, has been investigated by, for example, \citet{Eyring1931, PloosterGarvin1956} and \citet{Fettisetal1960}. The HBr$+$H abstraction and exchange reactions have been studied by \citet{PloosterGarvin1956}, and by \citet{WhiteThompson1974} whose channel-by-channel rates are consistent with the total rate from \citet{EndoGlass1976}. Based on Table~\ref{tab:reactions}, excluding other reactions, the competition between the Br$+$H$_{2}$ formation route and destruction via the HBr$+$H abstraction reaction strongly favours atomic Br. Thus gas-phase neutral-neutral chemistry is not expected to contribute to HBr formation unless temperatures of ${\sim}1000\,$K -- possible in hot cores, outflow shocks, and inner regions of protoplanetary disks -- are involved. 

Due to its low first ionization potential ($11.8\,$eV), Br is easily ionized and HBr can form in ion-neutral chemistry via the set of reactions (4)--(8) in Table~\ref{tab:reactions}. By analogy with F and \cl, reactions (4) to (8) should be fast, of order $10^{-10}-10^{-7}\,$cm$^{3}$s$^{-1}$ \citep{NeufeldWolfire2009}. However, as pointed out by \citet{MayhewSmith1990}, the Br$^{+}$+H$_{2}$ reaction is endothermic. We adopt a H$_{2}$ and \hbrp\ dissociation energy (E$_{\rm D}$) difference of $\approx 6200\,$K, estimated from the proton affinity (PA) and ionization potential (IP) of Br via $E_{\rm a}{=}$PA(Br)+IP(Br)-IP(H$_{2}$)-E$_{\rm D}$(H$_{2}$), suggested by D.\,Neufeld (private communication). The branching ratio of the dissociative recombination reactions (7) and (8) is unknown, but the dissociation energy of HBr ($\rm D_{0}{\approx}3.78\,$eV) is lower than that of H$_{2}$ ($4.48\,$eV), while those of H\cl\ and HF are similar and higher \citep[$\rm 4.43$ and $5.87\,$eV,][]{Darwent1970NIST}. The branching ratio into the HBr${+}$H channel may thus be lower than the $10\,$\% of the equivalent \cl\ reaction, which would lower the fraction of Br stored in HBr.  For the photoionization and -dissociation rates, we adopt order-of-magnitude numbers from the corresponding \cl\ and F reactions in \citet{NeufeldWolfire2009}.

The formation of HBr via H$+$Br collision requires a three-body interaction and thus is most efficient on grain surfaces \citep[e.g.][]{Reeetal2004}.

\begin{table*}[!ht]
\centering
\caption{Chemical reaction network for bromine.}
\begin{tabular}{ c l l | l l | l l }
\hline\hline
\#      &       $R_{1}$                 & $R_{2}$               & $P_{1}$                       & $P_{2}$ & $k(T)\,$[cm$^{3}$s$^{-1}$]    &       Reference       \\      
\hline
(1)     &       Br                              & H$_{2}$               & HBr                     & H             & $8.3\times10^{-11}\times\exp{(-8812\,K/T)}$           & \citet{Fettisetal1960}  \\
(2)     &       HBr                             & H                     & Br                              & H$_{2}$       & $8.9\times10^{-11}\times\exp{(-684\,K/T)}$    & \citet{WhiteThompson1974}       \\
(3)     &       HBr                             & H$^{\prime}$  & H$^{\prime}$Br        & H               & $4.0\times10^{-10}\times\exp{(-1140\,K/T)}$   & \citet{WhiteThompson1974}     \\
(4)     &       Br$^{+}$                        & H$_{2}$               & HBr$^{+}$               & H             & $10^{-9}\times\exp{(-6200\,K/T)}$     & \citet{MayhewSmith1990, NeufeldWolfire2009}$^{a,c}$     \\
(5)     &       HBr$^{+}$                       & e$^{-}$               & Br                              & H             & $2\times10^{-7}\times(T/300\,K)^{-0.5}$       & \citet{NeufeldWolfire2009}$^{a}$        \\
(6)     &       HBr$^{+}$               & H$_{2}$               & H$_{2}$Br$^{+}$         & H             & $(13.2\pm1.6)\times10^{-10}$  & \citet{BelikovSmith2008}      \\
(7)     &       H$_{2}$Br$^{+}$ & e$^{-}$               & HBr                   & H               & ${\leq}10^{-8}\times(T/300\,K)^{-0.85}$       & \citet{NeufeldWolfire2009}$^{a,b}$    \\
(8)     &       H$_{2}$Br$^{+}$ & e$^{-}$               & Br                            & 2H              &${\sim}10^{-7}\times(T/300\,K)^{-0.85}$        & \citet{NeufeldWolfire2009}$^{a}$      \\
(9)     &       HBr                             & $h\nu$                & Br                              & H             & $1.7\cdot 10^{-7}\times\chi_{\rm UV}$    & \citet{NeufeldWolfire2009}$^{a}$      \\
(10)    &       HBr                             & $h\nu$                & HBr$^{+}$               & e$^{-}$       & $10^{-10}\times\chi_{\rm UV}$ & \citet{NeufeldWolfire2009}$^{a}$        \\
\hline
\end{tabular}
\label{tab:reactions}\\
\flushleft
\emph{Notes. }$R_{i}$ and $P_{i}$ denote the reactants and products. $^{a}$ -- assumed order-of-magnitude similar to corresponding \cl, F reactions from \citet{NeufeldWolfire2009}; $^{b}$ -- upper limit based on H$_{2}$Cl$^{+}+e^{-}$ branching ratio (see text); $^{c}$ -- see Section~\ref{sec:chem}.
\end{table*}

\subsection{Chemical modelling results}
\begin{figure}[!ht]
\includegraphics[clip=,width=1.0\linewidth]{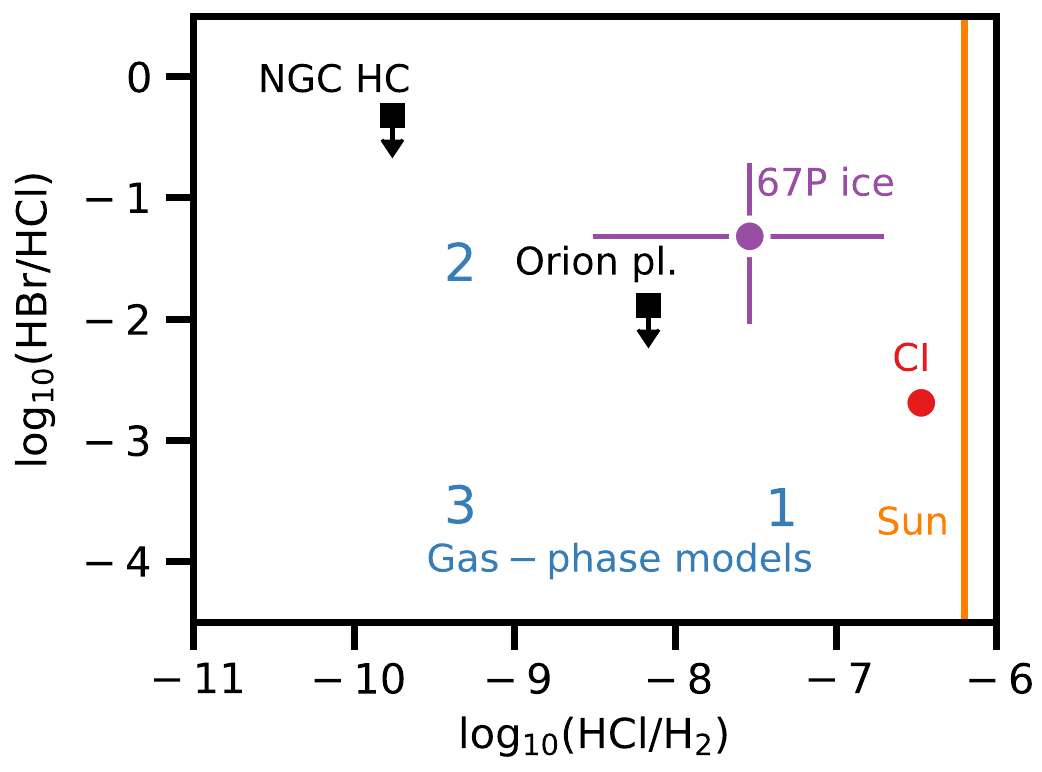}
\caption{The ratio of HBr to \hcl\ abundance in the NGC\,6334I Hot Core and the Orion\,KL~Plateau (upper limits for both sources) and pure gas-phase chemical models (1, 2, 3; see text), and in comet 67P/C-G \citep{Dhoogheetal2017}. We also show the elemental \cl/H$_{2}$ and Br/\cl\ ratios for meteorites (red circle) and the sun (orange line; the solar Br abundance is unknown). Models are shown for $n_{\rm H}=10^{6}\,$cm$^{-3}$ and $T_{\rm kin}=150\,$K. Variations of $\pm1$ in $\log{(n_{\rm H})}$ and $\pm50\,$K in $T_{\rm kin}$ induce negligible and $0.5\,$dex variations, respectively. Model 1 has all elemental \cl\ and Br in the gas; in 2 only \cl, and in 3 both \cl\ and Br are depleted from the gas by a factor of $100$.} 
\label{fig:chemmodel}
\end{figure}

We appended the reactions from Table~\ref{tab:reactions} to the OSU2009 network and ran time-dependent, gas-phase only simulations to $1.5\,$Myr with the \texttt{Astrochem} gas-phase chemistry code \citep{MaretBergin2015}. We ignored freeze-out in order to test the relevance of the \emph{Herschel} upper limits with the highest possible gas-phase elemental abundances. The physical conditions were set to $A_{\rm V}=20\,$mag (assuming a standard interstellar radiation field), $n_{\rm H}{=}10^{6}\,$cm$^{-3}$, and $T_{\rm kin}{=}150\,$K. The initial halogen abundances were either entirely atomic ions (\cl$^{+}$ and Br$^{+}$) or entirely diatomic hydrides (\hcl\ and HBr), but this had only a minor impact on the end-state abundances. We show the modelling results in Figure~\ref{fig:chemmodel} for three cases: 1) all elemental Br and \cl\ in the gas-phase; 2) undepleted Br and \cl\ depleted from the gas-phase by two orders of magnitude 3) both Br and \cl\ depleted by two orders of magnitude. Varying $n_{\rm H}$ by an order of magnitude had little impact on the abundances, while varying $T_{\rm kin}$ by $50\,$K induced a scatter of $0.5\,$dex in the plotted logarithmic abundance ratios. None of the models were strongly constrained by the upper limits, as we discuss below.

For the adopted physical conditions, the chemical network predictions place the gas-phase HBr abundance two orders of magnitude below the observed upper limit for the Orion~KL Hot Core. All literature studies of the gas-phase \cl\ abundance in protostellar sources find gas-phase \cl\ depletions of at least a factor $100$ to $1000$ \citep{Dalgarnoetal1974, Blakeetal1986, Schilkeetal1995, Pengetal2010, Kamaetal2015}. However, the ice fraction in the Orion~KL Hot Core is likely very small, so we expect model 1 to provide a reasonable prediction of the gas-phase (H)\cl\ and (H)Br abundance in this source.

In the NGC~6334I Hot Core, HBr may be just below the upper limit from \emph{Herschel} if elemental Br is not depleted from the gas, while \cl\ is known to be depleted by a factor $1000$. This seems unlikely.

\section{Conclusions}\label{sec:conclusions} 

We present the first search for bromine-bearing molecules in the interstellar medium, employing archival \emph{Herschel}/HIFI data. No detections of HBr or \hbrp\ are made, and we report upper limits of HBr for Orion~KL, Sgr~B2 (N), and NGC~6334I. Most of these upper limits lie above the values expected from a simple scaling down of \hcl\ emission using the \cl/Br elemental ratio.

In the Orion~KL Hot Core, the HBr/H$_{2}$O gas-phase abundance ratio is constrained to be an order of magnitude lower than the measured ratio in comet 67P/C-G. This result, along with the low HBr/CH$_{3}$OH ratio in all our sources and the low HBr/\hcl\ in the Orion~KL Plateau, is consistent with our chemical network modelling for Br, which predicts a low fraction of elemental Br in HBr in the gas phase. Our results suggest the HBr detected in high abundance in comet 67P/C-G formed in icy grain mantles.

\begin{acknowledgements}
The authors wish to thank Nathan Crockett for providing the data on Orion~KL and Sgr~B2, and Peter Schilke for the data on NGC6334I. We also thank Frederik Dhooghe for discussing his results on halogens in 67P/C-G and David Neufeld and Catherine Walsh for discussions on the chemistry. MK is supported by an Intra-European Marie Sklodowska-Curie Fellowship. Astrochemistry in Leiden is supported by the Netherlands Research School for Astronomy (NOVA), by a Royal Netherlands Academy of Arts and Sciences (KNAW) professor prize, and by the European Union A-ERC grant 291141 CHEMPLAN. HIFI has been designed and built by a consortium of institutes and university departments from across Europe, Canada and the United States under the leadership of SRON Netherlands Institute for Space Research, Groningen, The Netherlands and with major contributions from Germany, France and the US. Consortium members are: Canada: CSA, U.Waterloo; France: CESR, LAB, LERMA, IRAM; Germany: KOSMA, MPIfR, MPS; Ireland, NUI Maynooth; Italy: ASI, IFSI-INAF, Osservatorio Astrofisico di Arcetri-INAF; Netherlands: SRON, TUD; Poland: CAMK, CBK; Spain: Observatorio Astron\'{o}mico Nacional (IGN), Centro de Astrobiolog\'{i}a (CSIC-INTA). Sweden: Chalmers University of Technology - MC2, RSS \& GARD; Onsala Space Observatory; Swedish National Space Board, Stockholm University - Stockholm Observatory; Switzerland: ETH Zurich, FHNW; USA: Caltech, JPL, NHSC.
\end{acknowledgements}

\bibliographystyle{aa}
\bibliography{bromine}

\appendix

\section{Linelist of HBr transitions in range of HIFI}

The HIFI instrument on the \hso\ covered the three lowest rotational transition groups of HBr, which are summarised in Table~\ref{tab:hbrspec}. J = 1$_{x}$ $\rightarrow$ 0$_{2}$ fell in band 1a, J = 2$_{x}$ $\rightarrow$ 1$_{y}$ in band 4a and J = 3$_{x}$ $\rightarrow$ 2$_{y}$ in band 6a.

   \begin{table*}[!ht]
      \caption[]{H$^{79/81}$Br transitions between 500 and 1501 GHz.}
         \label{tab:hbrspec}
         \center
         \begin{tabular}{c c c c c c}
            \hline
            \hline
            \noalign{\smallskip}
             $\nu$ H$^{79}$Br & $\nu$ H$^{81}$Br & A    & E$\rm_{upper}$         & J',F'         & J'',F''  \\
                         MHz & MHz              & s$^{-1}$              & K                       &               &     \\
            \noalign{\smallskip}
            \hline
            \noalign{\smallskip}
            500540.1280 & 500407.2010 & 3.34E-4 & 24.0          & 1$_{1}$       & 0$_{2}$   \\      
            500647.7450 & 500497.3850 & 3.34E-4 & 24.0          & 1$_{3}$         & 0$_{2}$ \\      
            500780.0980 & 500607.7750   & 3.34E-4       & 24.0  & 1$_{2}$       & 0$_{2}$  \\
            1000859.5610 & 1000589.5640 & 5.33E-4       & 72.1  & 2$_{1}$       & 1$_{2}$ \\            
            1000993.2470 & 1000701.3110 & 1.71E-3       & 72.1  & 2$_{2}$       & 1$_{2}$ \\            
            1001089.1700 & 1000781.6850 & 2.24E-3       & 72.1  & 2$_{3}$       & 1$_{2}$ \\            
            1001089.1700 & 1000781.6850 & 3.20E-3       & 72.1  & 2$_{4}$       & 1$_{3}$ \\
            1001099.6240 & 1000790.3740 & 2.67E-3       & 72.1  & 2$_{1}$       & 1$_{1}$ \\
            1001125.5610 & 1000811.7800 & 1.60E-4       & 72.1  & 2$_{2}$       & 1$_{3}$ \\
            1001221.3420 & 1000891.7620 & 9.61E-4       & 72.1  & 2$_{3}$       & 1$_{3}$ \\
            1001233.1690 & 1000901.9040 & 1.33E-3       & 72.1  & 2$_{2}$       & 1$_{1}$ \\
            1500828.0700 & 1500397.4070 & 2.31E-4       & 144.1 & 3$_{2}$       & 2$_{3}$ \\
            1500923.7510 & 1500477.5790 & 3.24E-3       & 144.1 & 3$_{2}$       & 2$_{2}$ \\
            1500961.8100 & 1500509.4550 & 2.82E-3       & 144.1 & 3$_{3}$       & 2$_{3}$ \\
            1501025.2220 & 1500562.4860 & 9.91E-3       & 144.1 & 3$_{4}$       & 2$_{3}$ \\
            1501025.2220 & 1500562.4860 & 1.16E-2       & 144.1 & 3$_{5}$       & 2$_{4}$ \\
            1501057.6120 & 1500589.4380 & 8.10E-3       & 144.1 & 3$_{2}$       & 2$_{1}$ \\
            1501057.6120 & 1500589.4380 & 8.64E-3       & 144.1 & 3$_{3}$       & 2$_{2}$ \\
            1501094.0810 & 1500619.5480 & 1.10E-4       & 144.1 & 3$_{3}$       & 2$_{4}$ \\
            1501157.1420 & 1500672.5180 & 1.65E-3       & 144.1 & 3$_{4}$       & 2$_{4}$ \\
            \noalign{\smallskip}
            \hline
         \end{tabular} 
   \end{table*}

\clearpage

\section{Sgr B2(N) and NGC6334I at 500 GHz}\label{ap.sources}

\begin{figure}[!ht]
  \centering
        \label{Sgr_500_HBr}
    \includegraphics[clip=,width=1.0\columnwidth]{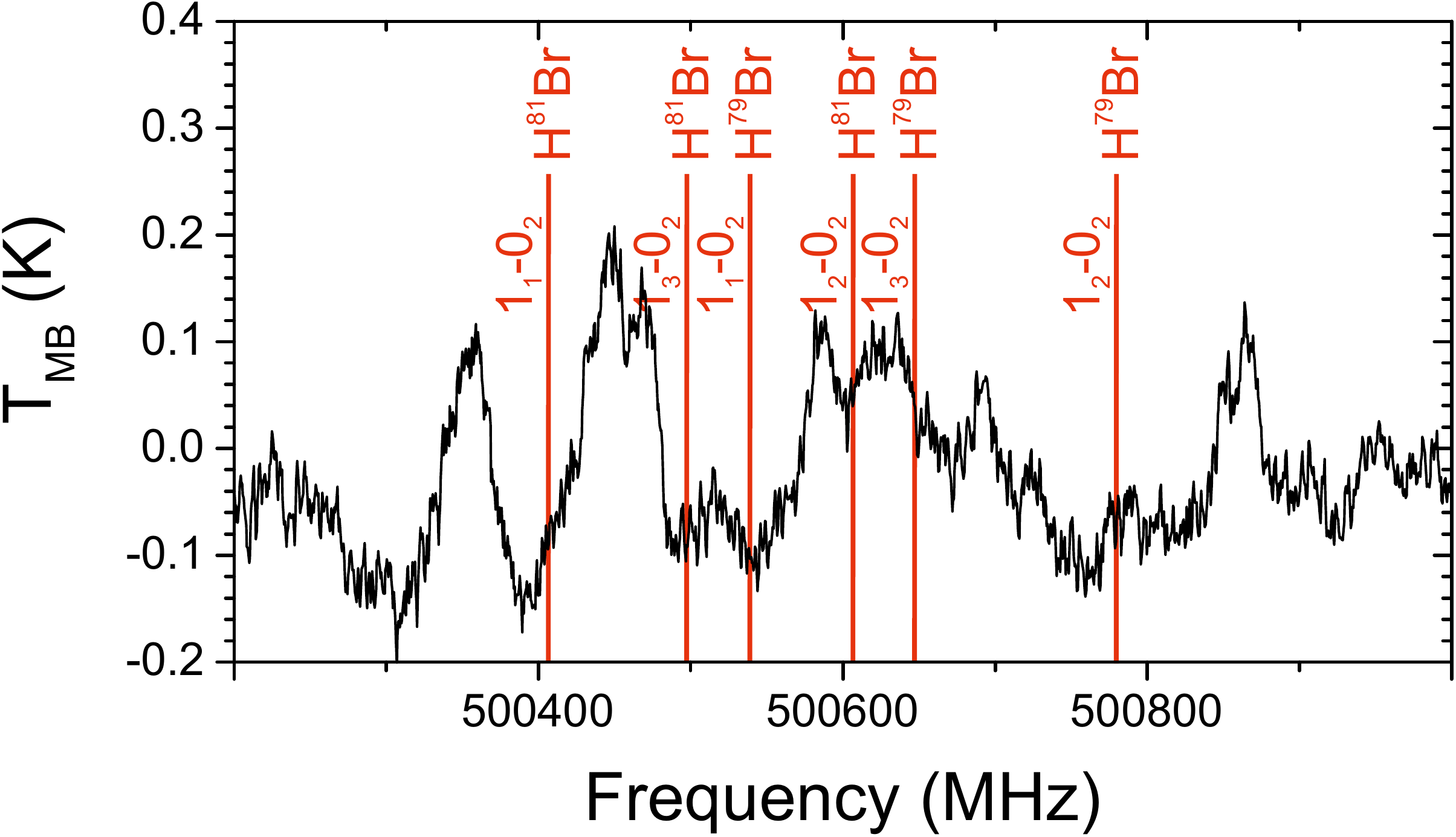}
    \caption{Positions of H$^{79/81}$Br transitions for J = 1$_{x}$ $\rightarrow$ 0$_{2}$ around 500 GHz in HIFI band 1a towards Sgr B2(N) for $V_{\rm LSR}$ = 64$\,$km$\,$s$^{-1}$.}
\end{figure}

\begin{figure}[!ht]
  \centering
        \label{NGC_500_HBr}
    \includegraphics[clip=,width=1.0\columnwidth]{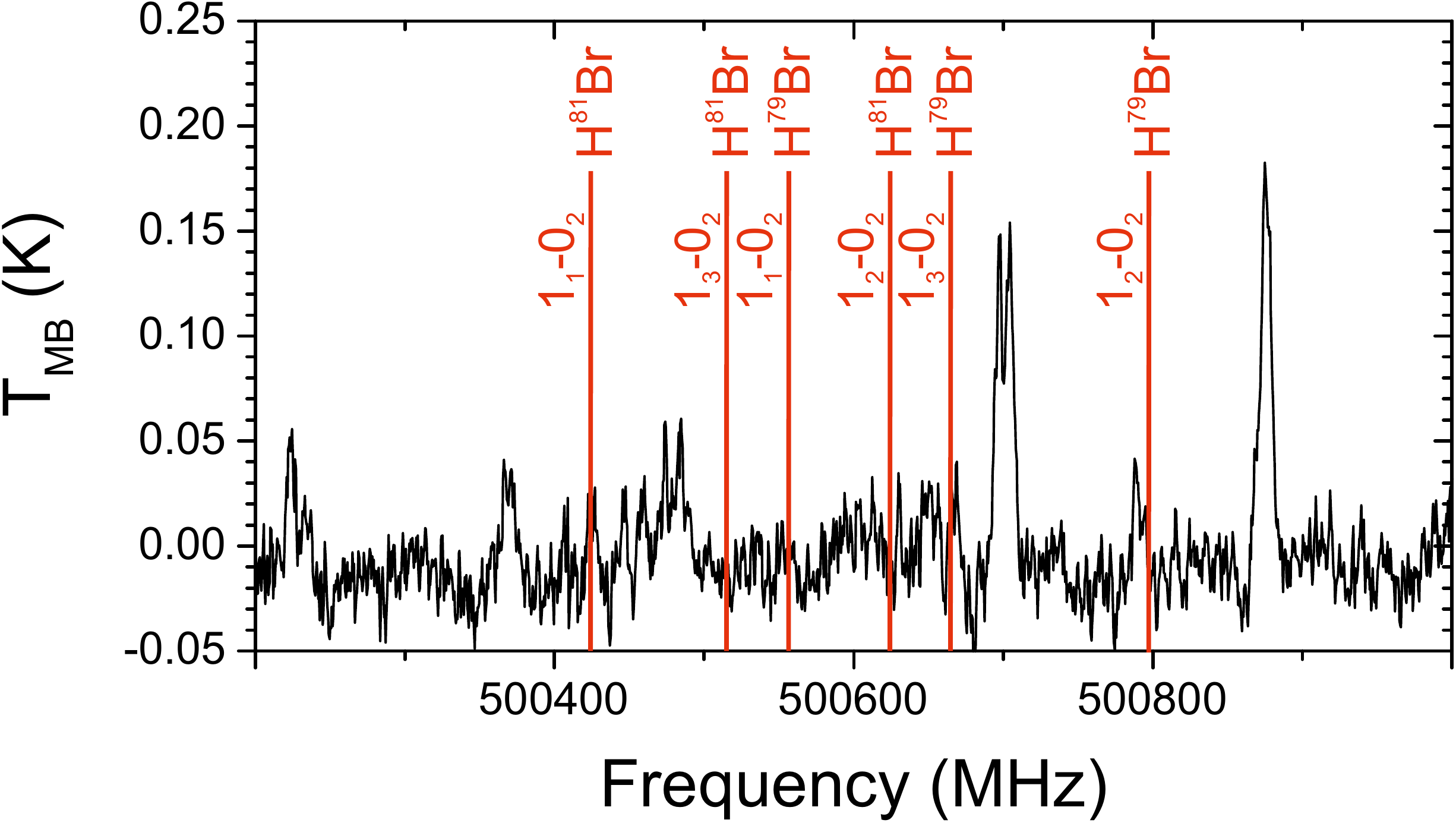}
    \caption{Positions of H$^{79/81}$Br transitions for J = 1$_{x}$ $\rightarrow$ 0$_{2}$ around 500 GHz in HIFI band 1a towards NGC6334I for $V_{\rm LSR}$ = -10$\,$km$\,$s$^{-1}$.}
\end{figure}


\section{Upper limit column densities of H$^{79+81}$Br and column densities of reference molecules}
\label{ap.uplim}

Table \ref{tab:hbrcoldens} lists the upper limit column densities of H$^{79+81}$Br for the full HIFI band 1a beam (= 44$^{\arcsec}$) in emission and absorption, calculated according to Eqs. \ref{eq.col_em} and \ref{eq.col_ab}. Upper limits have been derived for an excitation temperature of 100~K. The following columns in this table list the beam dilution correction factor and subsequently the beam dilution corrected upper limit column densities. 

Table \ref{tab:coldens} lists the column densities of the reference molecules H$_{2}$, H$_{2}$O, CH$_{3}$OH, HF, H$^{35+37}$\cl\ taken from \citet{Crockett2014b}, \citet{Neill2014} and \citet{Zernickel2012}.

\begin{table*}[t!]
      \caption[]{H$^{79+81}$Br column densities and beam dilution correction.}
         \label{tab:hbrcoldens}
         \centering
         \begin{tabular}{l c c c c c}
            \hline
            \hline
            \noalign{\smallskip}
            Source & \multicolumn{2}{c}{$N_{\rm T}$(H$^{79+81}$Br) (cm$^{-2}$)} & $\eta_{\rm BF}$ & \multicolumn{2}{c}{$N_{\rm S}$} \\
                        & Emission* & Absorption & & Emission* & Absorption  \\
            \hline
            \noalign{\smallskip}
            Orion KL Plat. & $\leq$7.9$\times$10$^{12}$ & $\leq$1.1$\times$10$^{14}$ & 3.2$\times$10$^{-1}$ & $\leq$2.5$\times$10$^{13}$ & $\leq$3.4$\times$10$^{14}$\\
            Orion KL HC  & $\leq$4.7$\times$10$^{12}$ & $\leq$3.8$\times$10$^{13}$ & 4.9$\times$10$^{-2}$ & $\leq$9.6$\times$10$^{13}$ & $\leq$7.8$\times$10$^{14}$\\
            Orion KL CR & $\leq$4.3$\times$10$^{12}$& $\leq$3.3$\times$10$^{13}$ & 4.9$\times$10$^{-2}$ & $\leq$8.8$\times$10$^{13}$ & $\leq$6.7$\times$10$^{14}$\\
            \noalign{\smallskip}
            Sgr B2(N) HC & $\leq$6.7$\times$10$^{12}$ & $\leq$4.0$\times$10$^{13}$ & 1.3$\times$10$^{-2}$ & $\leq$5.2$\times$10$^{14}$ & $\leq$3.1$\times$10$^{15}$\\
            Sgr B2(N) env. & $\leq$1.0$\times$10$^{13}$ & $\leq$8.9$\times$10$^{13}$ & -- & -- & -- \\
            \noalign{\smallskip}
            NGC6334I HC  & $\leq$1.2$\times$10$^{12}$ & $\leq$1.2$\times$10$^{13}$ & 1.3$\times$10$^{-2}$ & $\leq$9.2$\times$10$^{13}$ & $\leq$9.2$\times$10$^{14}$ \\
            NGC6334I env. & $\leq$1.7$\times$10$^{12}$ & $\leq$2.4$\times$10$^{13}$ & 1.7$\times$10$^{-1}$** & $\leq$1.0$\times$10$^{13}$ & $\leq$1.4$\times$10$^{14}$ \\
            \noalign{\smallskip}
            \hline
         \end{tabular} 
          
        \tablefoot{*Emission at $T_{\rm ex}$ = 100~K; **Beam dilution factor based on \hcl\ source size.}
\end{table*} 

\begin{table*}[t!]
      \caption[]{Column densities of the reference molecules H$_{2}$, H$_{2}$O, CH$_{3}$OH, HF, H$^{35+37}$\cl.\ }
         \label{tab:coldens}
         \centering
         \begin{tabular}{l c c c c c}
            \hline
            \hline
            \noalign{\smallskip}
            Source & H$_{2}$ & H$_{2}$O & CH$_{3}$OH & HF & H$^{35+37}$\cl\ \\
             & \multicolumn{5}{c}{(cm$^{-2}$)} \\
            \hline
            \noalign{\smallskip}
            Orion KL Plat.$^{a}$ & 2.8$\times$10$^{23}$ & 1.3$\times$10$^{18}$ & - & 2.9$\times$10$^{13}$* & 1.9$\times$10$^{15}$ \\
            Orion KL HC$^{a}$  & 3.1$\times$10$^{23}$ & 2$\times$10$^{20}$ & 6.8$\times$10$^{17}$  \\
            Orion KL CR$^{a}$ & 3.9$\times$10$^{23}$ & 1.8$\times$10$^{18}$ & 4.7$\times$10$^{17}$  \\
            \noalign{\smallskip}
            Sgr B2(N) HC$^{b}$ & 8$\times$10$^{24}$ & 5-10$\times$10$^{16}$ & 5$\times$10$^{18}$ \\
            Sgr B2(N) env.$^{b}$ & - & - & 1$\times$10$^{16}$ & 8.2$\times$10$^{14}$* & 1.3$\times$10$^{15}$* \\
            \noalign{\smallskip}
            NGC6334I HC$^{c}$  & 1.1$\times$10$^{24}$ & 2.1$\times$10$^{18}$ & 1.4$\times$10$^{19}$ & - & 1.9$\times$10$^{14}$\phantom{*} \\
            NGC6334I env.$^{c}$  & & & & 1.2$\times$10$^{14}$* & 1.5$\times$10$^{14}$*   \\
            \noalign{\smallskip}
            \hline
         \end{tabular} 
          
        \tablefoot{$^{a}$\citet{Crockett2014b}, $^{b}$\citet{Neill2014} $^{c}$\citet{Zernickel2012} and references therein; *absorption component.}
\end{table*} 

\begin{figure}[h!]
  \centering
    \includegraphics[width=0.5\textwidth]{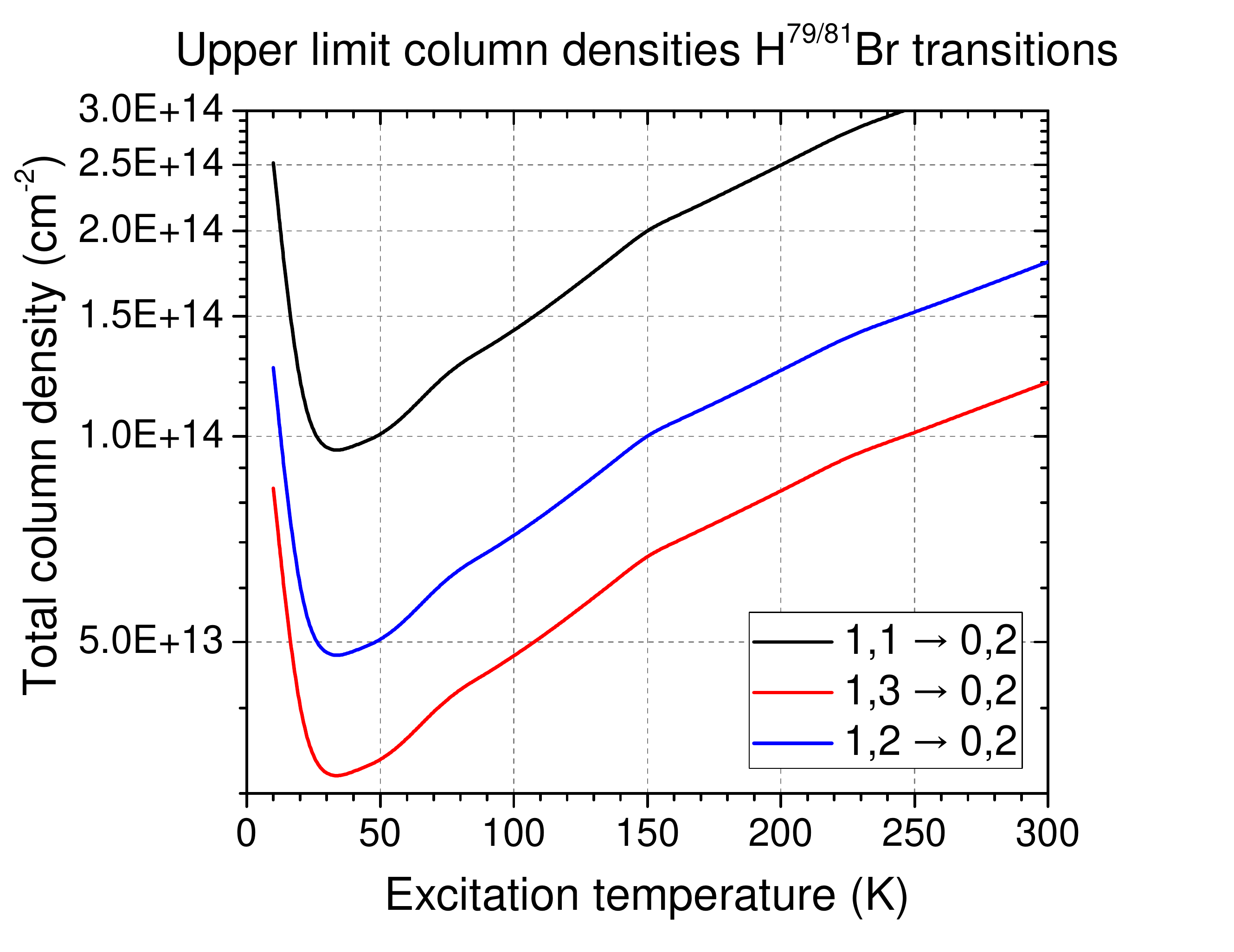}
    \caption{Upper limit column densities for the H$^{79/81}$Br (e.g. $^{79}$Br and $^{81}$Br are used interchangeably here) J = 1$_{x}$ $\rightarrow$ 0$_{2}$ transitions plotted versus rotational temperature based on the 3$\sigma$ values (216 mK km s$^{-1}$) found for the Orion KL Hot Core and beam-dilution corrected ($\eta$ = 0.049)}
    \label{fig:rotTplot}
\end{figure}

\end{document}